\documentclass[aps,twocolumn,prl,floatfix,showpacs,lengthcheck,citeautoscript,superscriptaddress,longbibliography]{revtex4-2}
\usepackage{amsmath}
\usepackage{amssymb}
\usepackage{graphicx}
\usepackage{bm}
\usepackage{color,soul}

\usepackage[dvipsnames,svgnames,table]{xcolor}
\usepackage{times}
\usepackage{MnSymbol}
\usepackage{bbold}
\usepackage[english]{babel}
\usepackage{hyperref}
\usepackage{orcidlink}

\hypersetup{
pdfstartview={FitH},
colorlinks=true,    
linkcolor=NavyBlue, 
citecolor=NavyBlue,   
filecolor=NavyBlue, 
urlcolor=NavyBlue   
}

\newcommand{\bea}{\begin{eqnarray}}
\newcommand{\eea}{\end{eqnarray}}
\newcommand{\be}{\begin{equation}}
\newcommand{\ee}{\end{equation}}

\definecolor{Nathanblue}{rgb}{0.94,0.317,0.5607}

\begin{document}

\title{Connecting the Many-Body Chern Number to Luttinger's Theorem through St\v{r}eda's Formula}

\author{Lucila {Peralta Gavensky}\,\orcidlink{0000-0002-5598-7303}}
\email{lucila.peralta.gavensky@ulb.be}
\affiliation{Center for Nonlinear Phenomena and Complex Systems, Universit\'e Libre de Bruxelles, CP 231, Campus Plaine, B-1050 Brussels, Belgium}
\author{Subir Sachdev\,\orcidlink{0000-0002-2432-7070}}
\affiliation{Department of Physics, Harvard University, Cambridge, Massachusetts 02138, USA}
\author{Nathan Goldman\,\orcidlink{0000-0002-0757-7289}}
\email{nathan.goldman@ulb.be}
\affiliation{Center for Nonlinear Phenomena and Complex Systems, Universit\'e Libre de Bruxelles, CP 231, Campus Plaine, B-1050 Brussels, Belgium}

\begin{abstract}
Relating the quantized Hall response of correlated insulators to many-body topological invariants is a key challenge in topological quantum matter. Here, we use St\v{r}eda's formula to derive an expression for the many-body Chern number in terms of the single-particle interacting Green's function and its derivative with respect to a magnetic field. In this approach, we find that this many-body topological invariant can be decomposed in terms of two contributions, $N_3[G] + \Delta N_3[G]$, where $N_3[G]$ is known as the Ishikawa-Matsuyama invariant and where the second term involves derivatives of Green's function and the self energy with respect to the magnetic perturbation. As a by-product, the invariant $N_3[G]$ is shown to stem from the derivative of Luttinger's theorem with respect to the probe magnetic field. These results reveal under which conditions the quantized Hall conductivity of correlated topological insulators is solely dictated by the invariant $N_3[G]$, providing new insight on the origin of fractionalization in strongly correlated topological phases.
\end{abstract}
\maketitle

\textit{Introduction.---}Exploring the interplay between strong correlations and topology is at the forefront of current research in condensed matter physics~\cite{Hohenadler2013,Rachel2018}. While topological invariants were originally defined within the framework of noninteracting band theory, where they are explicitly written in terms of one-body Bloch wave functions~\cite{Thouless1982,Hasan2010}, many-body generalizations have been proposed to characterize interacting quantum matter. An archetypal example is provided by the Hall conductivity of two-dimensional insulators, which is quantized according to $\sigma_{xy}\!=\!C e^2/h $,  where $C$ is known as the many-body Chern number:~a topological invariant constructed from the ground-state wave function and its derivatives with respect to twisted boundary conditions~\cite{Niu1985,Kudo2019}.
In the seminal works of Refs.~\cite{Volovik2009,Wang2010,Gurarie2011}, winding numbers of single-particle Green's functions defined in frequency-momentum space were also put forward as candidates to build up a topological classification of many-body phases of matter. In the case of a 2D quantum anomalous Hall insulator (QAHI), such a topological order parameter was identified as~\cite{Volovik2009,Gurarie2011,Essin2011,Wang2010,Wang2012}
\begin{eqnarray}
N_3[G]\!\!\!&=&\!\!\!\frac{\epsilon^{\lambda\nu\rho}}{24\pi^2}\!\!\int\!\!\!d^3 k e^{i\omega 0^{+}}\textrm{tr}\!\left[\! G_{\bm{k}}^{-1}\!\frac{\partial G_{\bm{k}}}{\partial k_{\lambda}} G_{\bm{k}}^{-1}\!\frac{\partial G_{\bm{k}}}{\partial k_\nu} G_{\bm{k}}^{-1}\frac{\partial G_{\bm{k}}}{\partial k_\rho}\! \right],\notag\\
 \label{N3} & &
\end{eqnarray}
where $G_{\bm{k}}(i\omega)$ is the continuous
imaginary frequency single-particle Green’s function (obtained in the limit of zero temperature), $k\!\!\!=\!\!\!(i\omega,k_x,k_y)$ is a three-momentum including the Matsubara frequency and $\epsilon^{\lambda\nu\rho}$ is the Levi-Civita tensor (a summation over indices is implicit); the trace is over the remaining internal degrees of freedom, such as orbital or spin variables. The invariant $N_3[G]$ is an integer associated with the third homotopy group of the general linear group of complex matrices $\pi_3(\textrm{GL}(\mathbb{C}))\!\cong\!\mathbb{Z}$, as provided by an invertible and single-valued Green's function defined in frequency-momentum space~\cite{Wu2021}~\footnote{In effective low-energy theories, the propagator of fractionalized quasiparticles can be multi-valued and lead to a fractional value of $N_3[G]$~\cite{Wu2021}.}.
In the noninteracting case, the integral over frequencies in Eq.~\eqref{N3} can be readily performed, yielding the first Chern number (or Thouless-Kohmoto-Nightingale–den Nijs invariant~\cite{Niu1985}), which quantifies the anomalous Hall conductivity of topological band insulators in two dimensions.

Broadly speaking, the Kubo-formula for the Hall conductance is explicitly written as a current-current correlation function~\cite{Mahan2000} so that, in principle, a four-point (two-body) Green's function is needed to account for the quantization of this transport coefficient in a correlated system~\cite{Yang2019}. Nevertheless, Ishikawa and Matsuyama have argued, based on a Ward-Takahashi identity, that $N_3[G]$ can still describe the quantized Hall response in the presence of electron-electron interactions~\cite{Ishikawa1986,Ishikawa1987,Imai1990}. This has been proved for certain tight-binding models in perturbation theory~\cite{Zhang2020}; see also Ref.~\cite{Blason2023} for a more recent derivation. Conversely, Eq.~\eqref{N3} was shown to fail capturing the Hall response in certain paradigmatic models of strongly interacting systems, such as fractional quantum Hall states~\cite{Gurarie2013} and topological Mott insulators~\cite{Zhao2023}.

Establishing the relationship between the topological invariants $C$ and $N_3[G]$, beyond the analysis of a particular model~\cite{Zhao2023}, is a formidable task in the context of correlated insulators:~Under which conditions are both quantities equal and how can one quantify their difference, remain as key open questions. In this Letter, we explicitly determine the difference between $C$ and $N_3[G]$ by making use of the Widom-St\v{r}eda formula~\cite{Streda1982,Widom1982,Streda1983}, which expresses $\sigma_{xy}$ as a density response to an external magnetic field. As a corollary, the topological invariants $C$ and $N_3[G]$ are shown to be equal whenever Luttinger's theorem holds.  This approach reveals a new interpretation of $N_3[G]$ in terms of the variation of Luttinger's theorem with respect to a probe magnetic field.

\textit{The Widom-St\v{r}eda formula.---} In an insulating state of matter, the Hall conductance can be expressed as a density response to a uniform external magnetic field $\textrm{B}$~\cite{Streda1982,Widom1982,Streda1983}
\begin{equation}
\label{Streda1}
\sigma_{xy} =  ec \frac{\partial n}{\partial \textrm{B}}\Bigg\rvert_{\mu, T=0},
\end{equation}
where $n$ is the particle density. This variation is performed at a fixed chemical potential $\mu$ lying within a spectral gap and in the limit of zero temperature $T=0$.  We can hence express the many-body Chern number as
\begin{equation}
\label{Streda}
C =  \phi_0\frac{\partial n}{\partial \textrm{B}}\Bigg\rvert_{\mu, T=0},
\end{equation}
where $\phi_0\!=\!hc/e$ stands for the flux quantum. The particle density of an interacting fermionic system at thermodynamic equilibrium is defined as~\cite{Mahan2000}
\begin{eqnarray}
\label{density}
n &=& \frac{1}{\beta \Omega}\sum_{n} e^{z_{n} 0^{+}}\textrm{Tr}\left[G(z_{n})\right],
\end{eqnarray}
where $G(z_{n})$ is the finite-temperature single-particle propagator. The fermionic Matsubara frequencies are given by $z_{n}\!=\!i\omega_{n}\!\!=\!i(2n+1)\pi/ \beta$ and $0^{+}$ represents an infinitesimally small positive number that assures convergence of the sum. Here $\Omega$ stands for the area of the system and the capital trace $\textrm{Tr}[\hdots]$ involves an integration over both internal and spatial degrees of freedom.  
Although St\v{r}eda's formula in Eq.~\eqref{Streda1} was originally derived in the noninteracting case within linear response theory~\cite{Streda1982}, it was also proven to be valid in generic interacting settings by making use of nonperturbative thermodynamic arguments~\cite{Widom1982,MacDonald1989}. 
Recently, fractional QAHIs in twisted bilayer samples were identified~\cite{Cai2023,Zeng2023} through this relation. It has also been particularly useful to detect the first realization of a bosonic fractional quantum Hall state in a cold-atom experiment~\cite{Leonard2023}, without relying on transport measurements~\cite{Repellin2020}.

\textit{Bloch Green's functions in a magnetic field.---}
We aim at formally evaluating the many-body Chern number of an anomalous quantum Hall state of matter from Eqs.~\eqref{Streda} and~\eqref{density}. In order to do so, we perform a perturbative expansion of the single-particle Green's function in the presence of a homogeneous magnetic field up to first order in $\mathbf{B}=\bm{\nabla}\times \textbf{A}$ by closely following the approach developed in Ref.~\cite{Kita2005}. The method builds upon the physical idea that, even though the discrete translational invariance of a periodic Hamiltonian is broken when including a vector potential $\textbf{A}(\bm{r})$ with a linear spatial dependence, any measurable quantity can be calculated in a way that explicitly preserves the original Bloch symmetry.
Dyson's equation of motion for $G(z_n)$ in a lattice reads
\begin{eqnarray}
\label{Dyson}
\sum_{\mathbf{R'}_{\!\!\!\alpha'}}\left[(z_n+\mu)\delta_{\mathbf{R}_{\alpha}^{}\mathbf{R'}_{\!\!\!\alpha'}} - H_{\mathbf{R}_{{\alpha}^{}}\mathbf{R'}_{\!\!\!\alpha'}}-\Sigma_{\mathbf{R}_{\alpha}^{}\mathbf{R'}_{\!\!\!\alpha'}}(z_n)\right]\\
\notag
\times\,\,G_{\mathbf{R'}_{\!\!\!\alpha'}\mathbf{R''}_{\!\!\!\alpha''}}(z_n)= \delta_{\mathbf{R}_{\alpha}^{}\mathbf{R''}_{\!\!\!\alpha''}},
\end{eqnarray}
where the coordinates $\mathbf{R}_{\alpha}$ denote the position of the atomic site $\alpha$ within the unit cell at $\mathbf{R}$ and the matrices $H$ and $\Sigma(z_n)$ are, respectively, the single-particle Hamiltonian and the self-energy. In the presence of a small magnetic field, the matrix elements of the single-particle Hamiltonian are modified as
\begin{equation}
H_{\mathbf{R}_{\alpha}^{}\mathbf{R'}_{\!\!\!\alpha'}} = H^{0}_{\mathbf{R}_{\alpha}^{}\mathbf{R'}_{\!\!\!\alpha'}} \textrm{exp}\left(i  \varphi_{\mathbf{R}_{\alpha}\mathbf{R'}_{\!\!\!\alpha'}}\right),
\end{equation}
where 
\begin{equation}
    \varphi_{\mathbf{R}_{\alpha}\mathbf{R'}_{\!\!\!\alpha'}} = \frac{2\pi}{\phi_0}\int_{\mathbf{R}_{\alpha}^{}}^{\mathbf{R'}_{\!\!\!\alpha'}}\textbf{A}(\bm{r})\cdot d\bm{r},
\end{equation}
is the Peierls phase factor~\cite{Peierls1933} and $H^{0}$ is the Hamiltonian in the absence of the field.
In the formalism of Ref.~\cite{Kita2005}, the gauge dependence of the single-particle Green's function and of the self-energy are factored out (see also~\cite{Luttinger1961,Laikhtmann1994,Khodas2003}), so that their matrix elements in coordinate space are expressed as
\begin{eqnarray}
\label{GB}
G_{\mathbf{R}^{}_{\alpha}\mathbf{R'}_{\!\!\!\alpha'}}(z_n) &=& \textrm{exp}\left(i  \varphi_{\mathbf{R}^{}_{\alpha}\mathbf{R'}_{\!\!\!\alpha'}}\right)G^{(\textrm{B})}_{\mathbf{R}^{}_{\alpha}\mathbf{R'}_{\!\!\!\alpha'}}(z_n),\\
\label{SigmaB}
\Sigma_{\mathbf{R}^{}_{\alpha}\mathbf{R'}_{\!\!\!\alpha'}}(z_n) &=& \textrm{exp}\left(i  \varphi_{\mathbf{R}^{}_{\alpha^{}}\mathbf{R'}_{\!\!\!\alpha'}}\right)\Sigma^{(\textrm{B})}_{\mathbf{R}^{}_{\alpha}\mathbf{R'}_{\!\!\!\alpha'}}(z_n).
\end{eqnarray}
Interestingly, the quantities identified by a superscript $(\textrm{B})$ depend only on the applied magnetic field and not on the arbitrary choice of gauge to describe the vector potential. Indeed, replacing Eqs.~\eqref{GB} and~\eqref{SigmaB} in Eq.~\eqref{Dyson} we find that they satisfy the modified Dyson's equation of motion
\begin{eqnarray}
\label{mod_Dyson}
\!\sum_{\mathbf{R'}_{\!\!\!\alpha'}}\!\!\Big\{\!\left[\!\left(z_n\!+\!\mu\right)\!\delta_{\mathbf{R}_{\alpha}^{}\mathbf{R'}_{\!\!\!\alpha'}}- H^{0}_{\mathbf{R}_{\alpha}^{}\mathbf{R'}_{\!\!\!\alpha'}}-\Sigma^{(\textrm{B})}_{\mathbf{R}_{\alpha}^{}\mathbf{R'}_{\!\!\!\alpha'}}(z_n)\!\right]\!\!\!\!& &\\
\notag
\!\!\times G^{(\textrm{B})}_{\mathbf{R'}_{\!\!\!\alpha'}\mathbf{R''}_{\!\!\!\alpha''}}\!(z_n)e^{i\frac{\pi}{\phi_0}\mathbf{B}\cdot (\mathbf{R'}_{\!\alpha'}^{}-\mathbf{R}^{}_{\alpha})\times (\mathbf{R''}_{\!\!\!\alpha''}-\mathbf{R'}_{\!\!\!\alpha '})}\!\Big\}\!\!\!\!&=&\!\!\!\delta_{\mathbf{R}_{\alpha}^{}\mathbf{R''}_{\!\!\!\alpha''}},
\end{eqnarray}
which is manifestly gauge and translationally invariant~\footnote{In this way, it is easy to verify that both the single-particle propagator and the self-energy in Eq.~\eqref{Dyson} get modified under a gauge transformation in the way originally established in Ref~\cite{Luttinger1961}.}. 

When Fourier transforming to $\bm{k}$-space, Eq.~\eqref{mod_Dyson} can be written to all orders in the magnetic field as~\cite{Kita2005}
\begin{eqnarray}
\notag
\sum_{\bm{k'}}\!\left\{\left[\!\left(\!z_n\!+\!\mu\!\right)\!\mathbb{1}\!-\!H^{0}_{\bm{k}+i \frac{\pi}{\phi_0} \mathbf{B}\times \bm{\nabla}_{\bm{k'}}}\!\!\!-\!\Sigma_{\bm{k}+i \frac{\pi}{\phi_0} \mathbf{B}\times \bm{\nabla}_{\bm{k'}}}^{(\textrm{B})}\!\right]\!G_{\bm{k'}}^{(\textrm{B})}\!\right\}\!\Bigg\rvert_{\bm{k}=\bm{k'}}\!\!\!\!\!\!\!\!\!\!\!\!&=&\\
 \notag
\!\!\sum_{\bm{k'}}\!\left\{\!e^{-i\frac{\pi}{\phi_0}\mathbf{B}\cdot(\bm{\nabla}_{\bm{k}}\times\bm{\nabla}_{\bm{k'}})}\!\left[\!\left(\!z_n\!+\!\mu\right)\!\mathbb{1}\!-\!H^{0}_{\bm{k}}\!-\!\Sigma_{\bm{k}}^{(\textrm{B})}\!\right]\!G_{\bm{k'}}^{(\textrm{B})}\!\right\}\!\Bigg\rvert_{\bm{k}=\bm{k'}}\!\!\!\!\!\!\!\!\!\!\!\!&=&\mathbb{1},\\
\label{Dyson_k}
& &
\end{eqnarray}
where the arguments $(z_n)$ have been omitted for the sake of brevity~\footnote{We remark that Ref.~\cite{Kita2005} provides a concrete prescription on how to diagrammatically compute $\Sigma^{(\textrm{B})}_{\bm{k}}$ by making use of the zero-field Feynman diagrams. Generically, the gauge-invariant self-energy has both an explicit field dependence and an implicit self-consistent one through $G^{(\textrm{B})}_{\bm{k}}$.}. Expanding both the self-energy and the exponential up to linear order in $\mathbf{B}$, we obtain
\begin{eqnarray}
\label{G-firstorder}
G^{(\textrm{B})}_{\bm{k}} &=& G_{\bm{k}}^{0}  + \frac{i \pi}{\phi_0}G_{\bm{k}}^{0}\mathbf{B}\cdot \left[\bm{\nabla}_{\bm{k}}(G_{\bm{k}}^{0})^{-1}\times \bm{\nabla}_{\bm{k}}G_{\bm{k}}^{0}\right]\\
\notag
& & +\,\mathbf{B}\cdot G^{0}_{\bm{k}}(\bm{\nabla}_{\mathbf{B}}\Sigma^{(\textrm{B})}_{\bm{k}})\Big\rvert_{\mathbf{B}=0}G^{0}_{\bm{k}} + \mathcal{O}(\textrm{B}^2),
\end{eqnarray}
where 
\begin{equation}
    G_{\bm{k}}^{0}(z_n) = \left[(z_n+\mu)\mathbb{1} - H_{\bm{k}}^{0}-\Sigma_{\bm{k}}^{0}(z_n)\right]^{-1},
    \label{G0}
\end{equation}
is the fully dressed two-point Green's function in the absence of magnetic field and $\Sigma_{\bm{k}}^{0}(z_n)$ its corresponding self-energy. 
An equivalent form of Eqs.~\eqref{Dyson_k} and~\eqref{G-firstorder} has already been obtained in Ref.~\cite{Onoda2006} for systems with continuous translational symmetry by introducing the Wigner representation in the Dyson's equation of motion. In the derivation presented above, which is explicitly written in Bloch basis, the periodic potential is fully taken into account. We also note that the noninteracting version of Eq.~\eqref{G-firstorder} was derived in Ref.~\cite{Chen2011}. The gauge-dependent Green's function $G_{\mathbf{R}_{\alpha}\mathbf{R'}_{\!\!\!\alpha'}}$, up to first order in $\mathbf{B}$, is readily obtained as the inverse Fourier transform of Eq.~\eqref{G-firstorder} multiplied by the phase factor $\textrm{exp}\left(i\varphi_{\mathbf{R}_{\alpha}\mathbf{R'}_{\!\!\!\alpha'}}\right)$, as prescribed by Eq.~\eqref{GB}.

\textit{Luttinger's theorem and the many-body St\v{r}eda response.---}
We are now ready to evaluate the many-body St\v{r}eda response [Eq.~\eqref{Streda}]. In order to do so, it will be instructive to rewrite the particle density in Eq.~\eqref{density} as
\begin{equation}
    n = n_1 + n_2,
\label{lutt_decomp}
\end{equation}
where we, respectively, defined Luttinger's density (also known as Luttinger's count~\cite{Heath2020}) and Luttinger's integral~\cite{Curtin2018} as
\begin{eqnarray}
\label{LC}
n_1 &=& -\frac{1}{\beta\Omega}\sum_n e^{z_n 0^{+}}\textrm{Tr}\left[G^{-1}(z_n)\frac{\partial G(z_n)}{\partial z_n}\right],\\
\label{LI}
    n_2 &=&  \frac{1}{\beta \Omega}\sum_n e^{z_n 0^{+}}\textrm{Tr}\left[\frac{\partial \Sigma(z_n)}{\partial z_n} G(z_n)\right].   
\end{eqnarray}
Equation~\eqref{lutt_decomp} is an exact decomposition, which directly follows from the Dyson's equation of motion of $G(z_n)$~\cite{Luttinger1960,Abrikosov1964}. Luttinger's theorem~\cite{Luttinger1960}, a cornerstone in the theory of many-body physics, states that
\begin{eqnarray}
    \lim_{T\to 0} n = \lim_{T\to 0} n_1,
    \label{Luttinger_theorem}
\end{eqnarray}
a result obtained by requiring a vanishing Luttinger's integral at zero temperature. This is trivially true for noninteracting systems ($\Sigma\!=\!0$), but is expected to hold for Fermi liquids~\cite{Oshikawa2000} and for generic correlated systems described by a Luttinger-Ward functional that reflects the $U(1)$ symmetry associated with electron-number conservation~\cite{Luttinger1960,Abrikosov1964,Sachdev2023}. As pointed out in Ref.~\cite{Seki2017}, Luttinger's count at zero temperature is related to a topological invariant, since it is directly proportional to the winding number in frequency space $N_1[G]$, namely,
\begin{equation}
\label{nL_N1}
    \lim_{T \to 0} n_1 = \frac{1}{\Omega}N_1[G],
\end{equation}
with
\begin{equation}
\label{N1}
    N_1[G] = -\frac{1}{2\pi i}\!\!\int\!\!dz e^{z 0^{+}}\textrm{Tr}\left[G^{-1}(z)\frac{\partial G(z)}{\partial z}\right].
\end{equation}

Using the definition of $G^{-1}$, implicitly given in Eq.~\eqref{Dyson}, and Eqs.~\eqref{GB} and~\eqref{SigmaB}, we can express Eqs.~\eqref{LC} and~\eqref{LI} in terms of the core or gauge-invariant propagator and self-energy as
\begin{eqnarray}
\notag
n_1\!\!\!&=&\!\!\!-\frac{1}{\beta}\!\sum_n\!\!\!\int\!\!\!\!\frac{d^2 k}{(2\pi)^2}e^{z_n 0^{+}}\!\textrm{tr}\!\left\{\!\left[\!(z_n\!+\!\mu)\mathbb{1}\!-\!H_{\bm{k}}^0\!-\!\Sigma_{\bm{k}}^{(\textrm{B})}\!\right]\!\frac{\partial G^{(\textrm{B})}_{\bm{k}}}{\partial z_n}\!\right\}\\
n_2\!\!\!&=&\!\!\! \frac{1}{\beta}\!\sum_n\!\!\!\int\!\!\!\!\frac{d^2 k}{(2\pi)^2}e^{z_n 0^{+}}\!\textrm{tr}\left[\!\frac{\partial \Sigma_{\bm{k}}^{(\textrm{B})}}{\partial z_n}G^{(\textrm{B})}_{\bm{k}}(z_n)\!\right],
\end{eqnarray}
where $\textrm{tr}[\hdots]$ implies summing only over internal degrees of freedom.  By considering Eqs.~\eqref{G-firstorder} and~\eqref{G0}, we explicitly obtain 
\begin{eqnarray}
\notag
    \!\!\!\!\phi_0\frac{\partial n_1}{\partial \textrm{B}}\Bigg\rvert_{\mu,\textrm{B}=0}\!\!\!\!\!\!\!\!\!\!\!\!\!\!\!&=&\!\!\!-i\frac{\pi}{\beta}\!\!\sum_n\!\!\!\int\!\!\!\!\frac{d^2 k}{\!(2\pi)^2\!}\epsilon^{0\nu\rho}\!e^{z_n 0^{+}}\!\!\!\textrm{tr}\!\left[\!G_{\bm{k}}^{0 -1}\!\frac{\partial G^{0}_{\bm{k}}}{\!\partial z_n \!}\!\frac{\partial G_{\bm{k}}^{0 -1}\!}{\!\partial k_\nu\!}\!\frac{\partial G^{0}_{\bm{k}}}{\!\partial k_\rho\!}\!\right]\!\!\!\!\!\!\\
    &-&\frac{\phi_0}{\beta}\!\sum_n\!\!\!\int\!\!\!\!\frac{d^2 k}{(2\pi)^2}e^{z_n 0^{+}}\frac{\partial}{\partial z_n}\textrm{tr}\Bigg[\!\frac{\partial \Sigma^{(\textrm{B})}_{\bm{k}}}{\partial \textrm{B}} G_{\bm{k}}^{0}\!\Bigg]\Bigg\rvert_{\mu,\textrm{B}=0}\!\!\!\!\!\!\!\!\!\!\!\!.
    \label{dBnL_T}
\end{eqnarray}
When taking the zero-temperature limit, the sum over Matsubara frequencies is replaced by a continuous integration along the imaginary frequency axis $1/\beta \sum_n \rightarrow \int dz/2\pi i$. In this regime, we can readily find from Eq.~\eqref{nL_N1} and the integration of Eq.~\eqref{dBnL_T} in the complex plane that the response of Luttinger's count to the external magnetic field is explicitly given by
\begin{eqnarray}
\label{dNL_dB}
\!\!\!\!\!\!\!\!\!\!\!\!\lim_{\substack{T\to 0\\ B\to 0}}\!\phi_0\frac{\partial n_1}{\partial \textrm{B}}\!\Bigg\rvert_{\mu}\!\!\!\!\!\!&=&\!\!\!\frac{\phi_0}{\Omega}\frac{\partial N_1[G]}{\partial \textrm{B}}\Bigg\rvert_{\mu,\textrm{B}=0}\!\!\!\!\!\!\\
\notag
\hspace{1cm}&=&\!\!\!\!N_3[G^{0}]\!+\!\frac{\phi_0}{\Omega\pi}\!\textrm{Im}\textrm{Tr}\!\left[\!\!\frac{\partial \Sigma^{(\textrm{B})}\!(\!i0^{+}\!)}{\partial \textrm{B}}G^0(\!i0^{+}\!)\!\!\right]\!\Bigg\rvert_{\mu,\textrm{B}=0}\!\!\!\!\!\!\!\!\!\!\!\!\!\!,
\end{eqnarray}
where $N_3[G^0]$ solely depends on the zero-field Green's function and is defined as in Eq.~\eqref{N3}. Here we have used that this winding number can be alternatively written in the more compact form
\begin{eqnarray}
\notag
N_3[G^0]\!\!&=&\!\!-\frac{\epsilon^{0\nu\rho}}{8\pi^2}\!\!\!\int\!\!\!d^2 k dz e^{z 0^{+}}\textrm{tr}\left[\!G^{0-1}_{\bm{k}}\frac{\partial G^{0}_{\bm{k}}}{\partial z}\frac{\partial G^{0 -1}_{\bm{k}}}{\partial k_{\nu}}\frac{\partial G^{0}_{\bm{k}}}{\partial k_{\rho}}\!\right],\\    
& &
\end{eqnarray}
which follows from the trace properties and $G_{\bm{k}}^{0 -1}\partial_{k_{\nu}}G^0_{\bm{k}} = -\partial_{k_{\nu}}G_{\bm{k}}^{0 -1} G^0_{\bm{k}}$. The last term in Eq.~\eqref{dNL_dB} is a contact (or boundary) term, stemming from the integral in frequency of the total derivative in Eq.~\eqref{dBnL_T}. It only depends on the analytical properties of the self-energy and Green's function at zero frequency. In an insulating state of matter with no poles of the gauge-invariant self-energy at the Fermi level, this term  exactly vanishes~\footnote{Interestingly, one can see that the boundary term in Eq.~\eqref{dNL_dB} should also vanish for a Landau-Fermi liquid. While $N_3[G]$ is not quantized in that case (due to the presence of a Fermi surface), it is still remarkable that it can be obtained from $N_1[G]$ through St\v{r}eda's response.}. Equation~\eqref{dNL_dB} is an exact result, and it is remarkable as it shows that a higher-order winding number of the fully dressed propagator can be obtained from $N_1[G]$ by taking its derivative with respect to a magnetic field~\cite{Prodan2016}.

On the other hand, we obtain
\begin{eqnarray}
\notag
\lim_{\substack{T\to 0\\ B\to 0}}\phi_0\frac{\partial n_2}{\partial \textrm{B}}\Bigg\rvert_{\mu}\!\!\!\!\!&=&\frac{\phi_0}{\Omega}\!\!\!\int\!\!\!\frac{dz}{2\pi i}e^{z 0^{+}}\!\!\frac{\partial}{\partial \textrm{B}}\textrm{Tr}\left[\!\frac{\partial \Sigma_{}^{(\textrm{B})}}{\partial z} G_{}^{(\textrm{B})}\!\right]\Bigg\rvert_{\mu,\textrm{B}=0}.\\
& &\label{dIL_dB}
\end{eqnarray}
Combining Eqs.~\eqref{dNL_dB} and~\eqref{dIL_dB} to compute the total density variation with respect to the probe magnetic field, we obtain that the many-body Chern number in Eq.~\eqref{Streda} is given by a sum of two contributions
\begin{equation}
C = N_3[G^0] + \Delta N_3[G],   
  \label{CvsN3}
\end{equation}
with
\begin{eqnarray}
\notag
\Delta N_3[G] &=& \frac{\phi_0}{\Omega\pi}\textrm{Im}\textrm{Tr}\left[\!\frac{\partial \Sigma^{(\textrm{B})}}{\partial \textrm{B}}(\!i0^{+}\!)G^0(\!i0^{+}\!)\!\right]\Bigg\rvert_{\mu,\textrm{B}=0}\!\!\!\!\\
\label{corrections}
&+&\frac{\phi_0}{\Omega}\!\!\!\int\!\!\!\frac{dz}{2\pi i}e^{z 0^{+}}\!\!\frac{\partial}{\partial \textrm{B}}\textrm{Tr}\left[\!\frac{\partial \Sigma_{}^{(\textrm{B})}}{\partial z} G_{}^{(\textrm{B})}\!\right]\Bigg\rvert_{\mu,\textrm{B}=0}\!\!\!\!\!\!\!\!\!\!\!\!.
\end{eqnarray}
Equation (25) constitutes one of the main results of the present work:~it offers an explicit expression for the difference between $C$ and $N_3[G^0]$ without relying on any model nor perturbative statement. We stress that, since it was derived from the Widom-St\v{r}eda formula, all the quantities are evaluated at a chemical potential $\mu$ within a spectral gap.\\

\textit{Analysis on some relevant cases.---} \textbf{i.}  In the absence of interparticle interactions ($\Sigma^{(\textrm{B})}=0$), $\Delta N_3[G]=0$ strictly vanishes and we recover the well-known result $C=N_3[G^0]$. A mean-field treatment leading to trivial self-energies independent of both $\omega$ and $\bm{k}$, i.e. that neglects spatial and temporal correlations, would also lead to $\Delta N_3[G]=0$~\footnote{A $\bm{k}$-independent self-energy $\Sigma^{(\textrm{B})}_{\bm{k}}\!=\!\Sigma^{(\textrm{B})}$ cannot have an explicit linear dependence on the magnetic field, such that $\partial \Sigma^{(\textrm{B})}/\partial \textrm{B}\!=\!0$.}.
\newline

\textbf{ii.}  In the particular case where $\partial\Sigma^{(\textrm{B})}/\partial \textrm{B}\rvert_{\textrm{B}=0}= 0$, we can explicitly quantify $\Delta N_3$ in terms of the zero-field Green's function and self-energy as
 \begin{equation}
\Delta N_3[G^0]\!=\!\frac{\epsilon^{0\nu\rho}}{8\pi^2}\!\!\!\int\!\!\!d^2 k\!\!\int\!\!dz \textrm{tr}\left[\!\frac{\partial \Sigma_{\bm{k}}^0}{\partial z}G_{\bm{k}}^0 \frac{\partial G_{\bm{k}}^{0 -1}}{\partial k_\nu}\frac{\partial G^{0}_{\bm{k}}}{\partial k_\rho}\!\right].
 \end{equation}
 The equation above is particularly relevant in the context of dynamical mean-field theory approaches, in which the self-energy is approximated as local but preserves its full frequency dependence. 
\newline

\textbf{iii.}  In the presence of interparticle interactions, we can further analyze the fate of $\Delta N_3[G]$ by considering the existence of the Luttinger-Ward functional $\Phi[G]$, which 
is differentially defined as~\cite{Luttinger1960,Potthoff2006,Kita2005,Abrikosov1964}
\begin{eqnarray}
\notag
\delta\Phi &=& \frac{1}{\beta}\sum_n \textrm{Tr}\left[\Sigma(z_n)\delta G(z_n)\right]\\
\label{dif_Phi}
 &=& \frac{1}{\beta}\sum_n \textrm{Tr}\left[\Sigma^{(\textrm{B})}(z_n)\delta G^{(\textrm{B})}(z_n)\right].     
\end{eqnarray}
The second equality holds on account of the cancellation of the Peierls phase factors when performing the trace operation. This was noted in Ref.~\cite{Kita2005}, where the functional was perturbatively written as an infinite sum over two-particle irreducible skeleton diagrams involving only the gauge-invariant propagator, namely
\begin{equation}
\label{skeleton}
\Phi\!=\!\frac{1}{\beta}\sum_n\sum_{l=1}^{\infty}\frac{1}{2l}e^{z_n 0^{+}}\textrm{Tr}\left[\!\Sigma^{(\textrm{B})}_l(z_n) G^{(\textrm{B})}(z_n)\!\right],  
\end{equation}
where $\Sigma^{(\textrm{B})}_l$ is the contribution of the $l$-th order diagram to $\Sigma^{(\textrm{B})}$.
Equations~\eqref{dif_Phi} and~\eqref{skeleton} imply that the self-energy can be written as an exact differential of Green's function, 
\begin{equation}
 \Sigma^{(\textrm{B})} = \beta\frac{\delta \Phi[G^{(\textrm{B})}]}{\delta G^{(\textrm{B})}}.  
 \label{Sigma_LW}
\end{equation}
The invariance of the Luttinger-Ward functional under a global shift of the Matsubara frequencies in $\delta z = 2\pi i/\beta$, $\Phi[G(z_n + \delta z)]=\Phi[G(z_n)]$, guarantees that~\cite{Luttinger1960,Abrikosov1964,Dzyaloshinksii2003}
\begin{align}
\notag
0 &=\sum_n e^{z_n 0^{+}} \textrm{Tr}\left[\frac{\delta \Phi}{\delta G^{(\textrm{B})}}\frac{G^{(\textrm{B})}(z_n+\delta z)-G^{(\textrm{B})}(z_n)}{\delta z}\right]\\
\notag
&=\frac{1}{\beta}\sum_n e^{z_n 0^{+}} \textrm{Tr}\left[\Sigma^{(\textrm{B})}(z_n)\frac{G^{(\textrm{B})}(z_n+\delta z)-G^{(\textrm{B})}(z_n)}{\delta z}\right].\\
& &
\label{zero_LI}
\end{align}We emphasize that this invariance is connected to the $U(1)$ symmetry of the theory describing the interacting system. Making an integration by parts in the second term of Eq.~\eqref{corrections} we find that $\Delta N_3[G]$ can be alternatively written as
\begin{eqnarray}
\notag
\Delta N_3[G] &=& -\frac{\phi_0}{\Omega\pi}\textrm{Im}\textrm{Tr}\left[\!\frac{\partial G^{(\textrm{B})}}{\partial \textrm{B}}(\!i0^{+}\!)\Sigma^0(\!i0^{+}\!)\!\right]\!\Bigg\rvert_{\mu,\textrm{B}=0}\!\!\!\!\\
\label{corrections2}
&-&\frac{\phi_0}{\Omega}\!\!\!\int\!\!\!\frac{dz}{2\pi i}e^{z 0^{+}}\!\!\frac{\partial}{\partial \textrm{B}}\textrm{Tr}\left[\! \Sigma_{}^{(\textrm{B})}\frac{\partial G_{}^{(\textrm{B})}}{\partial z}\!\right]\!\Bigg\rvert_{\mu,\textrm{B}=0}
\end{eqnarray}
The second term on the right-hand side of Eq.~\eqref{corrections2} can be identified as the derivative of Eq.~\eqref{zero_LI} with respect to a magnetic field in the limit $\delta z\!\rightarrow \!0$, so that it should vanish in the zero-temperature limit~\cite{Luttinger1960,Abrikosov1964}. Note, however, that this relies on regularity properties of the kernel of the discrete sum in Eq.~\eqref{zero_LI} near zero frequency~\cite{Potthoff2006,Fabrizio2022,Skolimowski2022}. On the other hand, the first term in Eq.~\eqref{corrections2} vanishes whenever the insulating state does not exhibit zeros of Green's function at the Fermi level, also known as Luttinger's surface~\cite{Dzyaloshinksii2003,Skolimowski2022}. Whenever all these conditions are satisfied, the many-body Chern number $C$, and hence $\sigma_{xy}$, should necessarily be given by the integer $N_3[G^0]$.
\newline

\textbf{iv.} Based on the previous discussion, we conclude that any disconnection between the integer $N_3[G^0]$ and the many-body Chern number $C$ is necessarily associated with a failure of Luttinger's theorem~\cite{Heath2020}.  This violation has been reported in a variety of systems~\cite{Altshuler1998,Georges2001,Rosch2007,Dave2013,Curtin2018,Blesio2018,Zitko2021}. In several cases, it has been associated with the existence of zeros of Green's function (poles of the self-energy), an ubiquitous situation in Mott insulators~\cite{Rosch2007,Wagner2023}, which also signals the breakdown of perturbation theory. The disagreement $N_3[G^0]\!\ne\!C$ that was recently identified in strongly correlated Hatsugai-Kohmoto Mott insulators~\cite{Zhao2023} falls into this category.

\textit{Concluding remarks.---} We have formally evaluated the Widom-St\v{r}eda response of a correlated QAHI. As a first step, we have perturbatively derived the corrections of the fully dressed single-particle Bloch Green's function in the presence of an external probe magnetic field. This allowed us to compute the first-order responses of both Luttinger's density and Luttinger's integral in the presence of the field. We have found that the many-body Chern number is generically given by a sum of two contributions: the Ishikawa-Matsuyama invariant $N_3[G^0]$ [Eq.~\eqref{N3}] and a correction $\Delta N_3[G]$ [Eq.~\eqref{corrections} or~\eqref{corrections2}]. When Eq.~\eqref{Luttinger_theorem} is satisfied, and in the absence of self-energy poles at the Fermi level, the correction $\Delta N_3[G]$ strictly vanishes. In this case, the many-body Chern number is directly given by the derivative of Luttinger's count with respect to the probe magnetic field
\begin{equation}
 C = \phi_0 \frac{\partial N_1[G]}{\partial \textrm{B}}\Bigg\rvert_{\mu,\textrm{B}=0}\!\!=N_3[G^0], 
\end{equation}
providing an important, so far overlooked, connection between the validity of Luttinger's theorem and the quantization of the Hall conductance with $N_3[G^0]$. We have rigorously established the disconnection between the Ishikawa-Matsuyama invariant and the many-body Chern number in strongly correlated phases of matter where Luttinger's theorem is violated, signaling the breakdown of the adiabatic continuity between the noninteracting and fully interacting ground state. Although our calculations were done by considering a magnetic perturbation on a QAHI, these can be easily generalized to the conventional quantum Hall (Landau level) scenario. In that case, the perturbative expansion can still be done with the probe field, while the preexistence of a strong quantizing magnetic field can be taken into account by using a magnetic Brillouin zone to define the gauge-invariant propagator in quasimomentum space.

Our results raise intriguing questions, such as the contribution of $N_3[G^{0}]$ and $\Delta N_3[G]$ to the fractional Hall response of strongly correlated states. Importantly, our results indicate that the fractionalization of the many-body Chern number must be encoded in $\Delta N_3[G]$, as provided in Eqs.~\eqref{corrections} and~\eqref{corrections2}. A numerical evaluation of $\Delta N_3[G]$ would thus provide insight on the emergence of fractionalization in strongly correlated topological matter.
 
\textit{Acknowledgments.---}
We acknowledge stimulating exchanges with Michele Fabrizio, Adolfo Grushin, Bruno Mera, C\'ecile Repellin,  Gonzalo Usaj and Botao Wang. Work in Brussels is supported by the FRS-FNRS (Belgium), the ERC Grant LATIS and the EOS project CHEQS. S. S. acknowledges the support of the Solvay Institutes, within the framework of the Jacques Solvay International Chairs in Physics, and the U.S. National Science Foundation (Grant No. DMR-2245246).

\end{document}